\documentclass[prd]{revtex4}
\usepackage{amsmath}
\usepackage{amssymb}
\usepackage{amsfonts}
\usepackage[dvips]{graphicx}

\begin{document}

\title{Exact solutions in bouncing cosmology}

\author{Tomasz Stachowiak}
\affiliation{Astronomical Observatory, Jagiellonian University,
Orla 171, 30-244 Krak{\'o}w, Poland}
\author{Marek Szyd{\l}owski}
\affiliation{Astronomical Observatory, Jagiellonian University,
Orla 171, 30-244 Krak{\'o}w, Poland}
\affiliation{M. Kac Complex Systems Research Centre, Jagiellonian University,
Reymonta 4, 30-059 Krak{\'o}w, Poland}

\begin{abstract}
We discuss the effects of a (possibly) negative $(1+z)^6$ type contribution to the
Friedmann equation in a spatially flat universe.
No definite answer can be given as to the presence and 
magnitude of a particular mechanism, because any test using the general 
relation $H(z)$ is able to estimate only the total of all sources of such 
a term. That is why we describe four possibilities: 1) geometric effects of 
loop quantum cosmology, 2) braneworld cosmology, 3) metric-affine gravity, and
4) cosmology with spinning fluid. We find the exact solutions for the models 
with $\rho^2$ correction in terms of elementary functions, and show all 
evolutional paths on their phase plane. Instead of the initial singularity, 
the generic feature is now a bounce.
\end{abstract}

\maketitle

\section{Introduction}

Cosmology, mainly due to its modern astronomical observations, seems to have
been limited to what is nowadays dubbed the Cosmological Concordance Model
(CCM). In the observational cosmology this role is played by the cold dark 
matter model with the cosmological constant ($\Lambda$CDM model). It offers the 
simplest explanation of the current Universe filled with two components--dark 
matter and dark energy \cite{Padmanabhan:2002ji,Copeland:2006wr}. The former 
is in the form of non-relativistic dust matter contributing one third of the 
total energy density. The latter has negative pressure, violating the strong 
energy condition, and constitutes the remaining two thirds. 

Interpretations of observational data within this model lead to the conclusion
that we live in an accelerating universe which is almost spatially flat and low
density \cite{Perlmutter:1998np,Riess:1998cb,deBernardis:2000gy}. Although the
introduction of $\Lambda$ could be justified on purely geometrical ground, many
authors seek a physical source of a suitably behaving matter, mimicking such a
geometrical effect. In literature there are many cosmological models with 
dark energy which take part in the contest for the best description of the 
accelerating Universe \cite{Szydlowski:2006pz,Szydlowski:2006ay}.

On the other hand, the very early universe, when classical physics is no longer
sufficient, is not so well understood. It is believed that ``initial
conditions'' for the classical evolution could be obtained from different,
fundamental physics which dominates when the energy density is high enough, and
which turn negligible as the Universe expands. In other words, new physics is
welcome in the process of cosmological model building
\cite{Puetzfeld:2005af,Kamionkowski:2002pc}. In particular, one could hope
for avoiding the initial singularity with a bounce of loop quantum cosmology
\cite{Date:2004fj}.

What we consider here are the effects of theories which predict a $\rho^2$-type
modification to the Friedmann equation, which fits the above description.
Although such reduction is a big
simplification, we feel it is important to stress that such problems are
explicitly solvable, which is often not checked before applying numerical
studies in current works; and that it is also possible to find and classify all
types of evolutional scenarios which help to gain insight into more complicated
models.

Below, we outline some recently significant possibilities of non-classical
physics, and proceed to solve the resulting Friedmann equation in the
subsequent sections.

\subsection{Braneworlds}

The basic idea is that our observational universe is some four dimensional
surface (called a brane) embedded in a more dimensional bulk spacetime, in
which the gravitational field, but not the others, can freely propagate. The
universe is self accelerating due to an
additional term appearing in the Friedmann equation when constrained to the
brane (for a pedagogical introduction into extra dimensions cosmology see e.g.
\cite{Rizzo:2004kr,Csaki:2004ay,Lue:2005ya}).

Let us consider a higher dimensional cosmology in the Randall-Sundrum framework
\cite{Randall:1999ee,Randall:1999vf}. Then the Friedmann equation assumes the
following form \cite{Shtanov:2002mb,Chung:1999zs}
\begin{equation}
    H^2 = \frac{\Lambda_4}{3} + \frac{8\pi}{3M_p^2}\rho +
    \epsilon\left(\frac{4\pi}{3M_5^3}\right)^2\rho^2 + \frac{c}{a^4},
    \label{RS_friedmann}
\end{equation}
where $\Lambda_4$ is the four dimensional cosmological constant,
$\epsilon=\pm1$, and $c$ is an integration constant whose magnitude as well as
sign depend on the initial conditions. The fifth term (called dark
radiation if $c<0$) we put equal to 0 because the third term is of the main
interest in the present discussion. This $\rho^2$ contribution arises from the
imposition of a junction condition for the scale factor of the brane. Note that the
$\rho^2$ term would decay as rapidly as $a^{-6}$ in a matter dominated universe,
thus modifying the early evolution and not being significant later on.

Both negative and positive $\epsilon$ are possible because $\epsilon$
corresponds to the metric signature of the extra dimension~\cite{Shtanov:2002mb}.
However, the sign of $\epsilon$ is crucial because models with timelike extra
dimension can avoid initial singularity by the so called ``bounce''
\cite{Brown:2004cs}. Of course, the sign of $\epsilon$ remains an open
question -- in \cite{Sahni:2002vs}, the authors discuss some consequences of the choice
$\epsilon=-1$.
Then, the presence of the $\rho^2$ term on the right-hand side
of (\ref{RS_friedmann}) leads a contracting universe to a bounce instead of a
big bang type of curvature singularity ($\rho\rightarrow\infty$,
$R_{abcd}R^{abcd}\rightarrow\infty$). If that happens, we have a constraint on
the value of brane tension $|\sigma|\gtrsim (1\mathrm{MeV})^4$ because the
bounce takes place at densities greater than during nucleosynthesis.

It should also be added that string theory does not provide any reasons for an
additional timelike dimension. It is, nevertheless, a mathematically viable
possibility, that some authors choose to investigate.

Note that such modification was also recently investigated in the context of
two-brane model \cite{Cline:2002ht}, and such equations arise in classically
constrained gravity of Gabadadze and Shang \cite{Gabadadze:2005ch,McInnes:2005sa}.

\subsection{Loop quantum universes}

The standard Big Bang cosmology presents us with the problem of initial
conditions for the Universe. What we can observe from ``the inside''
cannot answer the question why particular conditions were chosen, as we are
simply looking at the dynamics. Unfortunately we cannot look from ``the
outside'' and risk any kind of statistical or other external explanation. Many
authors shift the problem to the Planckian epoch, during which the quantum
effects are important. A contraction phase preceding the present expansion, as
predicted by some of these theories, seems to be one of the more attractive
scenarios. It is very interesting that the geometric effects in Loop Quantum
Cosmology predict a $-\rho^2$ modification to the Friedmann equation
\cite{Singh:2005xg,Ashtekar:2006wn}. This modification is relevant in the very
early universe when its matter energy density becomes comparable with the
Planck density. The effective Friedmann equation becomes
\begin{equation}
    H^2 = \frac{8\pi G\rho}{3}\left(1-\frac{\rho}{\rho_{\rm crit}}\right) +
    \frac{\Lambda}{3},
\end{equation}
where $\rho_{\rm crit} = \sqrt3/(16\pi^2\gamma^2)\rho_{\rm pl}$
\cite{Singh:2006im}, and one can distinguish classical and quantum bounces
depending on the relative magnitude of $\rho$ and $\rho_{\rm crit}$.

This is an application of the LQG methods, where the spacetime is discrete on
the quantum level, considerably affecting the large scale
\cite{Bojowald:2006da,Bojowald:2001xe,Ashtekar:2003hd} with nonperturbative
corrections. The $\rho^2$ modification is shared with the braneworld model, and
some comparison work can be found in
\cite{Singh:2006sg,Piao:2004hr,Lidsey:2006md}, see also \cite{Singh:2003vx}.

\subsection{Non-Riemannian cosmologies}

This misleadingly dubbed concept does not in fact involve the abandonment of
Riemannian manifold structure, but the name has been widely used so far and we
also adopt it here.

The first non-Riemannian cosmological models were based on the Einstein-Cartan
theory, which is a modification of general relativity by adding the torsion of
the space-time \cite{Trautman:2006fp}. In such models, the effects of spin and
torsion are manifested by the presence of an additional $-\rho^2\propto -a^{-6}$
contribution in the Friedmann equation. The main motivation for studying such
models was the problem of the initial singularity which could be avoided due to
the spin effects.

Among various extensions of general relativity, the so called metric-affine
gravity (MAG) is recently of interest. In contrast to the Riemann-Cartan theory,
the connection is no longer metric which implies that the covariant derivative
of the metric does not vanish. For a review see e.g.
\cite{Puetzfeld:2004yg,Puetzfeld:2005af}.

This time, the Friedmann equation reads
\begin{equation}
    H^2 = \frac{8\pi G\rho}{3} +\frac{\Lambda}{3}+\nu\frac{\psi^2}{a^6},
\end{equation}
where the new constant $\nu$ can be of both signs, and $\psi$ is an integration
constant. Therefore, the non-Riemannian quantities of this model (torsion and
non-metricity) will modify very early stages of evolution and are negligible at
later times.

We find different possible interpretations of the presence of the $\rho^2$
contribution in the Friedmann equation. However, we must remember that
cosmography which is based on the behaviour of the null geodesics maps the
geometry and kinematics of the universe in terms of $H(z)$, without any
reference to the source of each contribution. This happens because it measures
only the average properties of the matter density, and eventually only the
overall term $(-)(1+z)^6$ is significant \cite{Bludman:2006cg}.

Interestingly, similar $a^{-6}$ modifications have also been obtained in
universes with varying constants \cite{Barrow:2004ad}. It was shown that in the
generic case, alpha varying models lead to a bouncing universe.

\section{Dynamics of the model}

We take the Friedmann equation in the following form
\begin{equation}
    \frac{a'(t)^2}{a(t)^2} = 
    \frac{8\pi G}{3}\rho(t) + \nu\rho(t)^2 + \frac{\Lambda}{3},
\end{equation}
with a general $\rho^2$ term, and a new, real constant $\nu$. As mentioned
before,
spatially flat space is assumed, although there exist curved models
in which the dynamics is also solvable by means of elementary functions
\cite{Setare:2004jx,Setare:2005be}. We define a scale factor $x$, so that it is
equal to unity at time $t_0$ (say, today), and accordingly, the scaling law for
the (dust) matter density
\begin{equation}
    a(t)=a_0x(t),\quad \rho(t)=\rho_{m0}x(t)^{-3}.
\end{equation}
The main equation is now
\begin{equation}
    \frac{x'(t)^2}{x(t)^2} =
    \frac{\Lambda}{3} + \frac{8\pi G\rho_{m0}}{3}x(t)^{-3} +
    \nu\rho_{m0}^2 x(t)^{-6}.
\end{equation}
Next, we introduce the density parameters $\Omega$
\begin{equation}
    \rho_{m0}=\frac{3H_0^2}{8\pi G}\Omega_{m0},\qquad
    \nu\rho_{m0}^2=H_0^2\Omega_q,\qquad
    \Lambda= 3 H_0^2\Omega_{\Lambda},
\end{equation}
which requires a change of the time variable
\begin{equation}
x(t)=y(H_0 t)=y(s).
\end{equation}
so that the equation is simplified to a well known form
\begin{equation}
    \frac{y'(s)^2}{y(s)^2}=
    \Omega_{\Lambda}+\Omega_{m,0}y(s)^{-3} +
    \Omega_q y(s)^{-6},
\end{equation}
with the condition binding the present values of densities
\begin{equation}
    1 = \Omega_{\Lambda}+\Omega_{m,0}+\Omega_q. \label{bind}
\end{equation}
Note, that in our notation these $\Omega$'s are all constants.
Finally, introducing a new dependent variable $u$
\begin{equation}
    y(s)=\sqrt[3]{u(s)}
\end{equation}
the right-hand side becomes a polynomial
\begin{equation}
    \frac19 u'(s)^2 = 
    \Omega_{\Lambda}u(s)^2 + \Omega_{m,0}u(s) + \Omega_q.
\end{equation}
which can immediately be solved by the use of trigonometric functions only as
\begin{equation}
    u(s) =
    \frac{-2\Omega_{m0}\sqrt{\Omega_{\Lambda}}+
    (\Omega_{\Lambda}+\Omega_{m,0}^2-4\Omega_{\Lambda}\Omega_q)
    \cosh[3\sqrt{\Omega_{\Lambda}}(\alpha\pm s)]+
    (\Omega_{\Lambda}-\Omega_{m,0}^2+4\Omega_{\Lambda}\Omega_q)
    \sinh[3\sqrt{\Omega_{\Lambda}}(\alpha\pm s)]
    }
    {4\Omega_{\Lambda}^{9/2}},
\end{equation}
where $\alpha$ is the integration constant. Obviously such a general solution
is not very clear, the more so because $\alpha$ is in general a complex number.
That is why we simplify the main equation a bit more with another set of
substitutions (provided that $\Lambda\neq 0$)
\begin{equation}
    u(s) = w(3\sqrt{|\Omega_{\Lambda}|}s) = w(\tau),\qquad
    \Omega_q=q|\Omega_{\Lambda}|,\quad
    \Omega_{m0} = \mu|\Omega_{\Lambda}|,
\end{equation}
and obtain
\begin{equation}
    w'(\tau)^2 = {\rm sign}(\Lambda) w(\tau)^2 + \mu w(\tau) + q, \label{main}
\end{equation}
or, if $\Lambda=0$,
\begin{equation}
    w'(\tau)^2 = \Omega_{m,0} w(\tau) + 1 - \Omega_{m,0}, \label{zero_lambda}
\end{equation}
upon using (\ref{bind}). Depending on the sign of $\Lambda$, we have three main
groups of solutions, which we proceed to describe in more detail.

\section{$\Lambda>0$}

\begin{figure}[h]
\begin{center}
\includegraphics[width=0.7\textwidth]{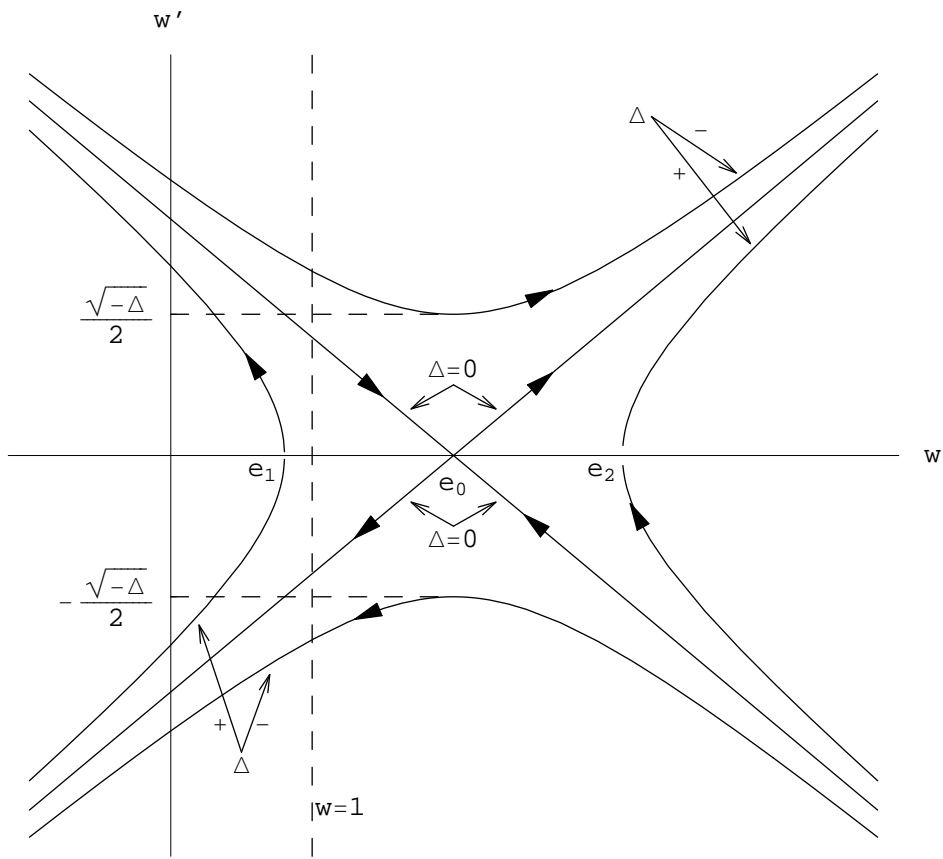}
\caption{The possible evolutional paths of the system when $\Lambda>0$,
depending on the sign of $\Delta=\mu^2-4q$. Only the paths intersected by the
$w=1$ line are admissible in a given set of the parameters. $e_0=-\mu/2$,
$e_{1,2}=(-\mu\mp\sqrt{\Delta})/2$. \label{hyper}}
\end{center}
\end{figure}

\begin{figure}[h]
\begin{center}
\includegraphics[width=0.7\textwidth]{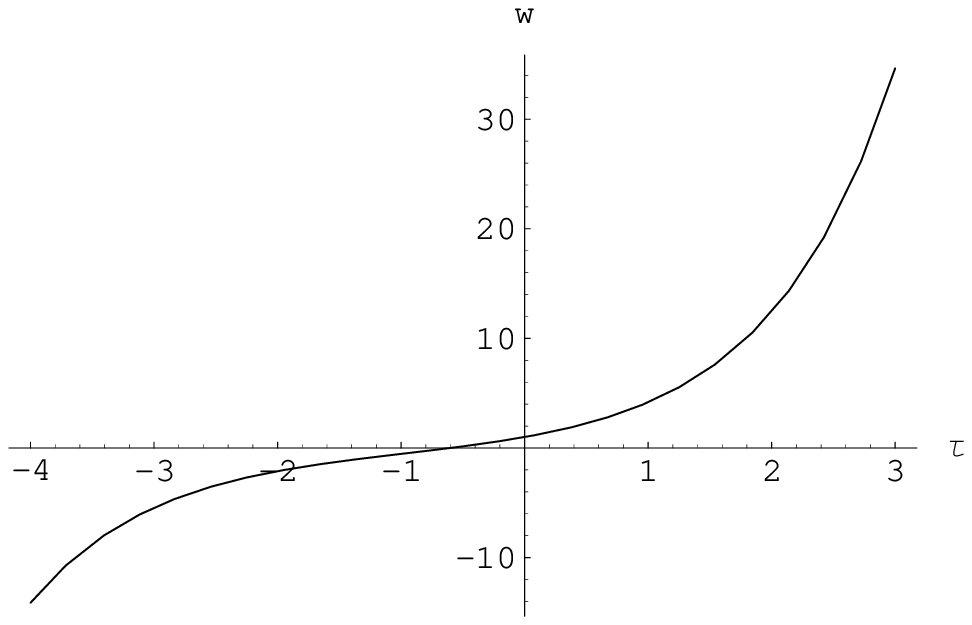}
\caption{$\Lambda>0$ and $\Delta<0$ solution. \label{hyper_mon}}
\end{center}
\end{figure}

The curves described by equation~(\ref{main}) in the plane $(w,w')$ are either
two hyperbolae or straight lines, as shown in Figure~(\ref{hyper}). If
$\Delta=\mu^2-4q$ is negative, we have the hyperbolae situated above and below
$e_0=-\mu/2$ - they are really the same solution, only with the sign of time
interchanged. The main equation always admits such a symmetry, but as we want a
universe which is expanding at this moment, we choose the solution with
$w'(0)>0$. It is given by
\begin{equation}
    w(\tau) =
    \frac12\left[-\mu+(2+\mu)\cosh(\tau)+2\sqrt{1+q+\mu}\sinh(\tau)\right],
    \label{sol_pos}
\end{equation}
and shown in figure (\ref{hyper_mon}).

\begin{figure}[h]
\begin{center}
\includegraphics[width=0.7\textwidth]{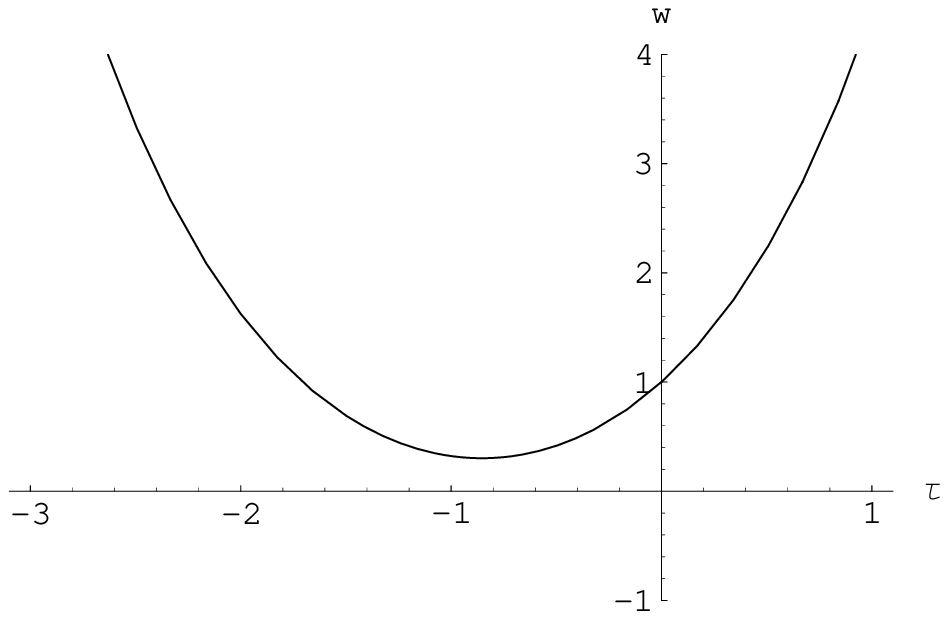}
\caption{$\Lambda>0$ and $\Delta>0$ solution. \label{hyper_bounce}}
\end{center}
\end{figure}

When $\Delta>0$, we have the other two hyperbolae, but as
$\Omega_{m,0}$ and $\Lambda$ are both positive, $e_0$ is negative, and so only
the right branch can intersect with the $w=1$ line. If
$e_2=(\mu+\sqrt{\Delta})/2 > 0$, a bounce without singularity appears.
The formula is the same as in
(\ref{sol_pos}), and is shown in figure (\ref{hyper_bounce}).

This is in fact
the generic setup, as $\Lambda$ is usually assumed to be positive, and the
$\rho^2$ contribution negative. This means that $q$ is negative, making
$\Delta$ positive. For this reason we consider bounce the generic feature here.

\begin{figure}[h]
\begin{center}
\includegraphics[width=0.7\textwidth]{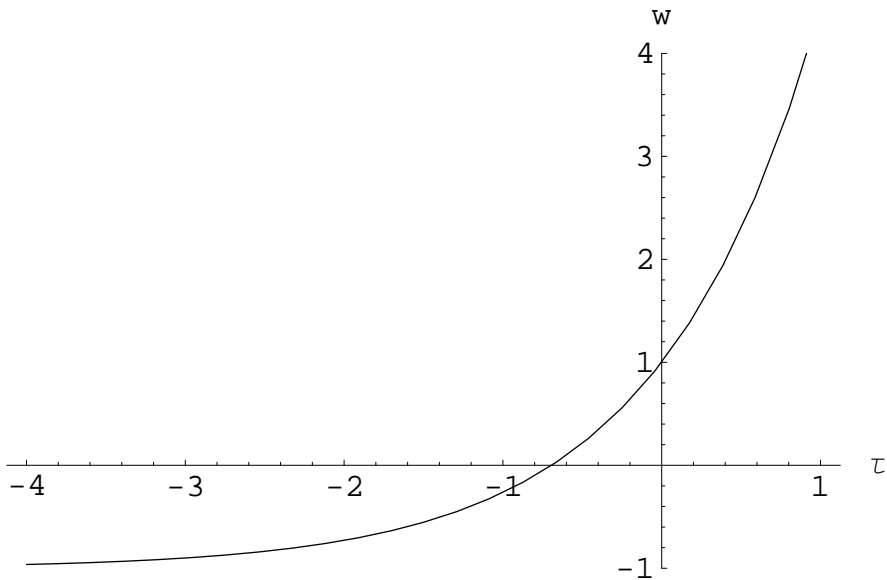}
\caption{$\Lambda>0$ and $\Delta=0$ solution. \label{hyper_asympt}}
\end{center}
\end{figure}

If $\Delta=0$, $e_0$ becomes a double root of the polynomial in the right-hand
side of equation (\ref{main}) and accordingly a possible static solution.
However, it is negative in our physical setup, so the only possibility are the
straight-line solutions. They are different in that they asymptotically evolve
from the static solution in the infinite past, as can be seen in 
figure~(\ref{hyper_asympt}). Formula~(\ref{sol_pos}) still holds.

\section{$\Lambda<0$}

\begin{figure}[h]
\begin{center}
\includegraphics[width=0.7\textwidth]{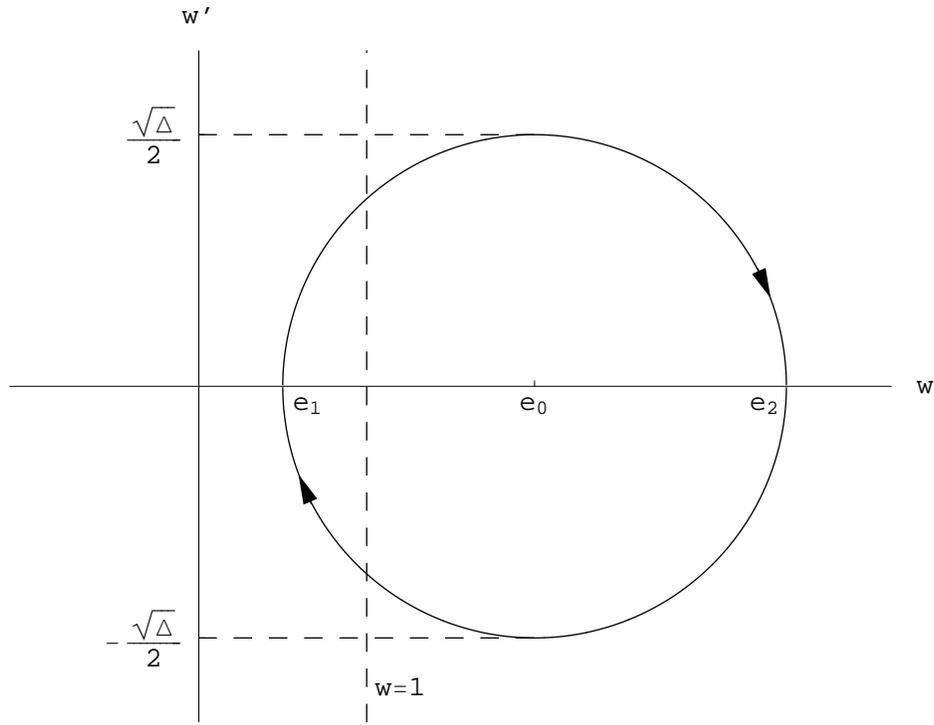}
\caption{The possible evolutional paths of the system when $\Lambda<0$.
The circle always intersects the $w=1$ line, unless $\Delta=0$, and the
solution reduces to the static one $w=e_0=-\mu/2$.
$e_{1,2}=(-\mu\mp\sqrt{\Delta})/2$. \label{circle}}
\end{center}
\end{figure}

\begin{figure}[h]
\begin{center}
\includegraphics[width=0.7\textwidth]{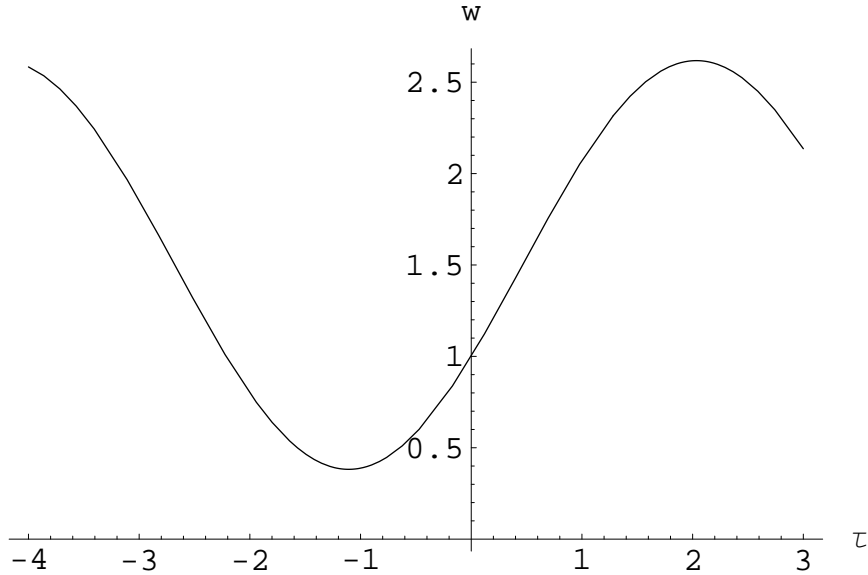}
\caption{$\Lambda<0$ and $\Delta>0$ solution. \label{circle_period}}
\end{center}
\end{figure}

The situation is much simpler now, as shown in figure~(\ref{circle}) --- the
only possible curve is a circle when $\Delta=\mu^2-4q$ is positive. The
condition (\ref{bind}) means that the circle must intersect the $w=1$ line,
unless $\Delta=0$, which corresponds to a stable, static solution. If
$e_1=(-\mu-\sqrt{\Delta})/2$ is positive, the evolution is a singularity free
oscillation. Obviously only one formula is needed to describe this case
\begin{equation}
    w(\tau) = 
    \frac12\left[-\mu+(2+\mu)\cos(\tau)+2\sqrt{-1-q-\mu}\sin(\tau)\right],
\end{equation}
and a typical example is shown in figure (\ref{circle_period}).

\section{$\Lambda=0$}

Finally, the simplest case with zero cosmological constant, where going back to
equation (\ref{zero_lambda}). This time the solution is even simpler -
polynomial in time
\begin{equation}
    u(s)=1+3s+\frac94\Omega_{m,0}s^2,
\end{equation}
where, as before, we determined the integration constants by assuming $u(0)=1$
and $u'(0)>0$. This is also a bounce, with a minimum size of
\begin{equation}
    u_{\rm min} = 1-\frac{1}{\Omega_{m,0}},
\end{equation}
which is only positive if $\Omega_{m,0}>1$, or equivalently $\Omega_q<0$,
corresponding to a negative $\rho^2$ term in the Friedmann equation.
Qualitatively the behaviour here is the same as that in figure
(\ref{hyper_bounce}), only the expansion is not exponentially fast.

\section{Conclusions}

We have studied the FRW models with an additional $(1+z)^6$ term in the
Friedmann equation. Astronomical measurements based on the relation $H(z)$ and
general dynamics of such models ignore the number and the particular type of
contributions -- as long as they scale the same way. Among these,
we described: brane theory, non-Riemannian
modifications and loop quantum cosmology. Their combined effect, as observed
through the Hubble's relation, could in theory be null, due to the
indefiniteness of sign. Which is not to say that they would be
indistinguishable (or null) when other observations come into account.

Assuming the non-zero term $a^{-6}$, it is still possible to find exact
solutions, and classify all possible evolutional paths. It was shown that exact
solutions can be expressed in terms of elementary functions and depending on
the sign of $\Lambda$ we have three main families (with the exact formulas
provided in the respective sections).

For $\Lambda=0$ (section V) we obtain a simple algebraic solution which is a
bounce, and for a negative $\rho^2$ contribution the minimal value of the scale
factor is positive. Qualitatively it is the same as the one presented in Figure
\ref{hyper_bounce}.

When $\Lambda<0$ (section IV), we have a periodic behaviour (Figure
\ref{circle_period}) of the scale factor, but as above, the minima are not
necessarily positive.

Finally, taking $\Lambda>0$ (section III), we obtain a universe which expand
exponentially, but which could have followed three different paths in the past.
It could have bounced (Figure \ref{hyper_bounce}), it could have been expanding
all the time from minus infinity (Figure \ref{hyper_mon}) or asymptotically
from the static solution (Figure \ref{hyper_asympt}). Of course, the last two
scenarios, would have to be considered as singular and expanding from the $a=0$
singularity.

The generic feature seems to be the replacement of the
initial singularity by a bounce. However, the effects might influence also the
late evolution if one considers a phantom fluid, and are a possibility of
avoiding the so called Big Rip singularity \cite{Sami:2006wj}.

As mentioned before, similar exact solutions of the Friedmann equation were
found for brane scenarios by Setare \cite{Setare:2004jx,Setare:2005be}. Such
solutions are among the
simplest possible, where the left hand side of the equation is a polynomial of
degree 2, or can be reduced to such a polynomial. Usually this can be done by
either introducing a new ``conformal'' time, or choosing a new scale factor
which some power of $a$. For polynomials of greater degrees such reduction is,
in general, impossible and elliptic functions have to be used
\cite{Dabrowski:2004hx}.

\section*{Acknowledgements}

The paper was supported by Marie Curie Host Fellowship MTKD-CT-2004-517186 
(COCOS).


\end{document}